# HIGHLY SCALABLE, PARALLEL AND DISTRIBUTED ADABOOST ALGORITHM USING LIGHT WEIGHT THREADS AND WEB SERVICES ON A NETWORK OF MULTI-CORE MACHINES

Munther Abualkibash, Ahmed ElSayed, Ausif Mahmood

Department Of Computer Science, University of Bridgeport, Bridgeport, CT, USA
mabualki@bridgeport.edu, aelsayed@bridgeport.edu,
mahmood@bridgeport.edu

## ABSTRACT

*AdaBoost is an important algorithm in machine learning and is being widely used in object detection. AdaBoost works by iteratively selecting the best amongst weak classifiers, and then combines several weak classifiers to obtain a strong classifier. Even though AdaBoost has proven to be very effective, its learning execution time can be quite large depending upon the application e.g., in face detection, the learning time can be several days. Due to its increasing use in computer vision applications, the learning time needs to be drastically reduced so that an adaptive near real time object detection system can be incorporated. In this paper, we develop a hybrid parallel and distributed AdaBoost algorithm that exploits the multiple cores in a CPU via light weight threads, and also uses multiple machines via a web service software architecture to achieve high scalability. We present a novel hierarchical web services based distributed architecture and achieve nearly linear speedup up to the number of processors available to us. In comparison with the previously published work, which used a single level master-slave parallel and distributed implementation [1] and only achieved a speedup of 2.66 on four nodes, we achieve a speedup of 95.1 on 31 workstations each having a quad-core processor, resulting in a learning time of only 4.8 seconds per feature.*

## 1. INTRODUCTION

One of the challenging research topics in object detection in recent years has been face detection. Many approaches to reliable face detection have been attempted e.g., [1-4]. A survey on face detection techniques presented in [3] classifies recent work into four major categories: First, knowledge-based methods where human knowledge is the basis for face determination. Second, feature invariant approaches where facial feature structures are easy to find even in difficult conditions, such as poor lighting. Third, template matching methods where storing faces is important for matching purposes. Lastly, appearance-based methods which rely on faces training set to be learned before being able to detect faces.

In this paper, we focus on the method used in the last category. Viola and Jones have used this method in [4] and their algorithm has proven to be very successful in real time face detection. For learning of the face detection classifier, Viola Jones' algorithm uses AdaBoost [5] on a training set of approximately 5000 faces and 10,000 non faces. Since the training set and the number of possible features used in learning a classifier is quite large, running time of AdaBoost can be very high and the time to complete training of one feature can be on the order of several minutes. Thus depending on the total number of features desired, the learning time of the algorithm can be several days. For example, to obtain a 200 feature classifier, it takes about one day on a modern workstation. Thus for applications, where we would like the object detection to be adaptive, this training time needs to be greatly reduced. An example where this is highly desirable is identifying a particular model of a car when it gets stolen, and the traffic cameras need to detect only that model. Since there are tens of thousands of make model and year of cars, waiting for several days until a classifier is ready would not be an option.

Parallel and distributed processing can be employed for speeding up of the AdaBoost algorithm. With the wide availability of multicore workstations and high speed network of workstations, we can take advantage of the computing power available. Software frameworks targeting the efficient use of multiple cores within a workstation have been recently developed for both Windows and Unix/Linux environments e.g., Task Parallel Library [6] allows creation of light weight threads where the cost to launch a thread is only 50 assembly language instructions as opposed to approximately 200 instructions for a regular thread. Further, web services frameworks based on scalable WS* standards are available where efficient distributed implementation can be successfully developed. In this paper, we develop a hybrid parallel as well as distributed implementation of the AdaBoost algorithm exploiting both the multiple cores in a machine via the Task Parallel Library, as well as multiple machines via a novel hierarchical web services based distributed implementation to achieve significant reduction in training time.

## 1.1. Related Work

A recent work at a parallel and distributed implementation of AdaBoost has been reported in [1]. Their system consists of a client computer and a server pool. The parallelization of AdaBoost is accomplished by using feature blocks. They employ four computing nodes in the distributed model and thus achieve a speedup of only 2.66.

Another parallel implementation of face detection has carried out on GPUs in [7] and have demonstrated the improvement in face detection time of the Haar feature based Adaboost algorithm to about 30 frames per second (FPS) using CUDA on the GeForce GTX 480 GPU. The main problem with their work is that they focused on the object detection time, not the classifier training time, so if the training dataset is updated, the training process will be repeated again. Our main focus in this paper is the parallelization of training of the Haar feature classifier to make training process near real-time such that any changes in the feature types or the training dataset can be handled in a short retraining time.

The reminder of this paper is organized as follows. In section 2, we provide a brief background on the AdaBoost algorithm before presenting our parallel and distributed approach in section 3. Section 4 provides the experimental results. Conclusions are presented in section 5.

## 2. ADABOOST ALGORITHM

One of the main contributions of Viola and Jones is the integral image [4]. The benefit of using integral image is to speed up the computation of rectangular features used in AdaBoost. We review the calculations in the integral image below and then describe the AdaBoost algorithm.

## 2.1. Integral Image

To get the integral image value on position x, y, the summation of the all pixels values located on top and to the left of x, y is taken. Figure 1 explains this concept.

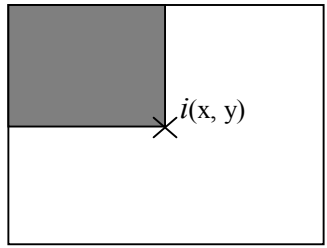

Figure 1. Summation of all pixels on top and to the left of x,y is the integral image value at x,y

The following equation explains the computations to get the integral image:

$$i(x,y) = \sum_{x' \le x, y' \le y} o(x', y')$$

where $i(x,y)$ represents the integral image, and $o(x,y)$ represents the original image.

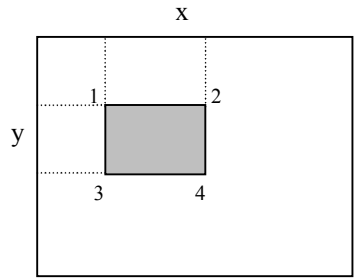

Figure 2. Calculating the integral image in a rectangular region.

For obtaining the integral image in the dark rectangle in Figure 2 only, the following equation is followed:

$$i(dark\ rectangle) = 4 + 1 - (2 + 3)$$

## 2.2. Extracted Feature Types and Selection

The features extraction in Viola Jones' algorithm are based on Haar basis functions [4, 8]. Five types of features have been used in the original algorithm as shown in Figure 3.

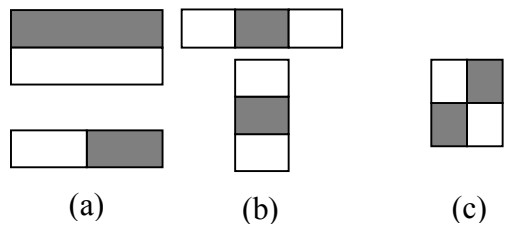

(a)          (b)          (c)

Figure 3. Five rectangular features. Figure (a) shows two rectangles horizontal and vertical features, figure (b) shows three rectangles horizontal and vertical features, and figure (c) shows a four rectangles feature.

To calculate the value of each one of the features, the sum of pixels located in the white side of the rectangle are subtracted from the dark side [4]. The window size used for training and detecting purposes in face detection is 24x24. For scaling, the starting point is the smallest size of a rectangular feature, e.g. in three rectangle feature type, it is 3x1 pixels, in a two rectangle feature type, it is 2x1 pixels, etc.. Each rectangular feature is scaled up until reaching total window size of 24x24. As a result, the total number of features for each type is:

- For three rectangle feature type, 27,600 features.
- For two rectangle feature type, 43,200 features.
- For four rectangle feature type, 20,736 features.

The total number for all features combined is going to be:

Three rectangles horizontal + three rectangles vertical + two rectangles horizontal + two rectangles vertical + four rectangles = 27600 + 27600 + 43200 + 43200 + 20736 = 162,336 features. During the learning phase, all of these features are going to be computed for all faces in the training set. The set of faces which we use for training purpose is the same one that has been used by Viola and Jones for face detection [4]. The size of each image is 24x24. There are 4916 faces and 7960 non-faces. Thus, the total number of all possible features in all training images is 2,090,238,336 (i.e., number of training images multiplied by features per image).

Viola and Jones have used AdaBoost to combine weak classifiers into a stronger classifier [5]. The conventional AdaBoost algorithm works by assigning good features relatively higher weight and the poor ones a smaller weight to determine the best weak classifier [4]. The main goal of a weak classifier is to get the optimal threshold among positive and negative examples for any rectangular feature. This technique has been known as the decision stump. The selected threshold minimizes the number of misclassified examples. The decision of a weak classifier is 1 or 0, i.e., positive or negative. The following equation explains the way a weak classifier works.

$$h(x, f, p, \theta) = \begin{cases} 1, & if \ pf(x) < p\theta \\ 0, & otherwise \end{cases}$$

where p is either 1 or -1, $\theta$ is the threshold, and f is the feature.

## 2.3. AdaBoost Algorithm used by Viola and Jones

- Suppose there are N numbers of images as training set. Each image is labelled as 0 for negative images, and, 1 for positive images, as shown in table 1.

Table 1. Example of images and labels.

| Images | $x_1$ | $x_2$ | $x_3$ | ........ | $x_N$ |
|--------|-------|-------|-------|----------|-------|
| Label | 1 | 0 | 0 | ........ | 1 |

- Initializing the weight for each image in the first round as shown in the following table:

Table 2. Example of images and labels and weights.

| Images | $w_{1,x_1}$ | $w_{1,x_2}$ | $w_{1,x_3}$ | ........ | $w_{1,x_n}$ |
|--------|-------------|-------------|-------------|----------|-------------|
| Label  | 1           | 0           | 0           | ........ | 1           |
| Weight | $\frac{1}{2l}$ | $\frac{1}{2m}$ | $\frac{1}{2m}$ | ........ | $\frac{1}{2l}$ |

Where l is the total number of positive images i.e., faces, and m is the total number of nonfaces.

- For t = 1 to T:

  1. Normalizing the weight of each image in each round as the following:
  $w_{t,x_i} = \frac{w_{1,x_i}}{w_{sum}}$ , where $w_{sum}$ is the sum total of the weight of all images in the same round.

  2. Calculating the error of all features, until getting the feature has the minimum error. The selected feature is the best weak classifier in $t^{th}$ round.

  $\epsilon_t = \min_{f,p,\theta} \sum_{x_i} w_{x_i} |h(x_i, f, p, \theta) - y_{x_i}|$

  3. Based on the minimum error ($\epsilon_t$), which is determined by $f_t$, $p_t$, and $\theta_t$, get $h_t(x)$, where:
  $h_t(x) = h(x, f_t, p_t, \theta_t)$ .

  4. As preparation for the next round, the weight should be updated:
  $$w_{t+1,x_i} = w_{t,x_i} \beta^{1-e_{x_i}}$$
  where $e_{x_i} = 1$, if $x_i$ is misclassified, $e_{x_i} = 0$ otherwise, and $\beta_t = \frac{\epsilon_t}{1-\epsilon_t}$ .

- At the end, after going through all rounds, the strong classifier is determined as the following:

$$C(x) = \begin{cases} 1, & \sum_{t=1}^{T} \alpha_t h_t(x) \geq \frac{1}{2} \sum_{t=1}^{T} \alpha_t \\ 0, & \qquad\qquad otherwise \end{cases}$$

where $\alpha_t = \log \frac{1}{\beta_t}$

# 3. OUR PARALLEL AND DISTRIBUTED WEB SERVICES- BASED ADABOOST ARCHITECTURE

The original AdaBoost determines the best weak classifier in each round based on the minimum classification error. It needs to go through all features and determine which feature yields the minimum error. Since there are a large number of features, the execution time during the learning phase is high. Our parallel approach speeds up the execution time by efficiently parallelizing the AdaBoost algorithm. We implement a three way approach to get results in shortest possible time. We run main computational part of AdaBoost in parallel using Task parallel Library (TPL). Task parallel library is a built in library in Microsoft .NET framework. The advantage of using TPL is noticed in multi-core CPUs, where the declared parallel workload is automatically distributed between the different CPU cores by creating light weight threads called tasks [6].

To further improve the execution time of AdaBoost, we use web services to run parallel Adaboost on multiple workstations in a distributed manner. Our first level architecture does workload division based on the feature type. Since there are five feature types, we use five workstations at level 1. As shown in figure 4, total of six machines are used, i.e., one master and five slaves. To achieve further scalability, computation of each of the feature types is further expanded to a second level using web services. Figure 5 shows twenty one PC's being used, in a two-level hierarchy, Master, Sub-Master, and Slaves.

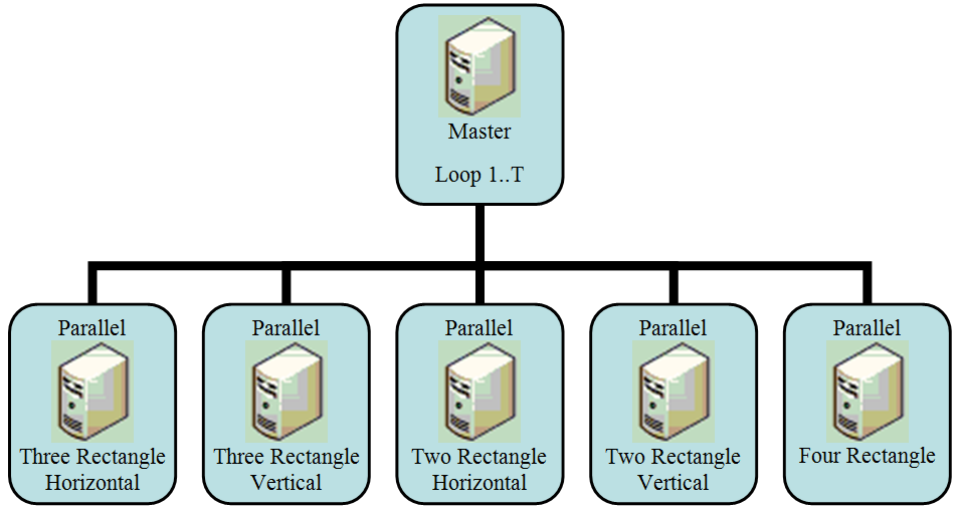

Figure 4. One hierarchal level for Web Services and Parallel AdaBoost, based on Master and five Salves (Total of six Pc's)

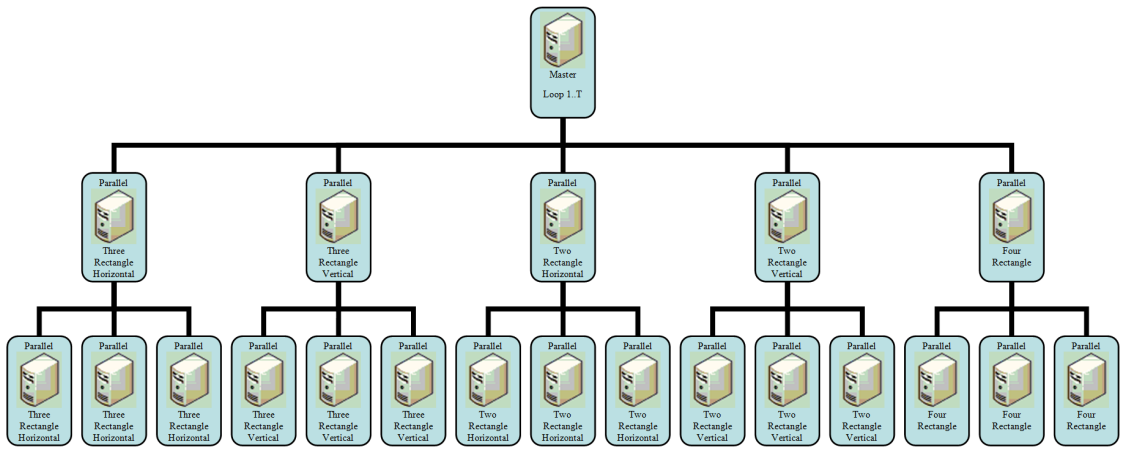

Figure 5. Two hierarchal level for Web Services and Parallel AdaBoost, based on Master and five Sub-Master and 3 Salves for each Sub-Master (Total of twenty one Pc's)

## 3.1. The Parallel AdaBoost Algorithm

Three approaches for speeding up execution are implemented in the AdaBoost algorithm:

- Parallel execution.
- Web Services and Parallel execution on one hierarchal level.
- Web Services and Parallel execution on two hierarchal levels.

### 3.3.1. Parallel execution

All features are grouped based on type, such as, three rectangle horizontal, three rectangle vertical, two rectangle horizontal, two rectangle vertical, and four rectangle. Each group is uploaded to the system memory in parallel. Once all of these have been loaded, then the AdaBoost rounds from 1 to T are started. Since the goal is to find the feature has the minimum error in each group in parallel in each round, five features are selected from five groups. Among them, the feature that has the least minimum error is picked and based on that the weight is updated for the next round of AdaBoost algorithm. Since selecting a minimum error feature runs in parallel, the execution time time is reduced by a factor of five.

### 3.3.2. Web Services and Parallel execution on one hierarchal level

Each group of features is distributed to a separate PC. Since five groups exist, five PC's are used for feature calculations, and a master coordinates the five PC's as shown in figure 4. The parallel and distributed pseudo code for this approach is described below.

### Pseudocode of one level master and five slaves Parallel Adaboost

- Given example images $(x_1, y_1), \ldots, (x_n, y_n)$ where $y_i = 0,1$ for negative and positive examples respectively.
- Prepare one master workstation and five slaves.
- Each slave is assigned to one particular feature type.
  - (slave 1, Three rectangles Horizontal)
  - (slave 2, Three rectangles Vertical)
  - (slave 3, Two rectangles Horizontal)
  - (slave 4, Two rectangles Vertical)
  - (slave 5, Four rectangles)
- On slave's workstations: Initialize all images on each slave.
- On master workstation:
  - Initialize weights $w_{1,i} = \frac{1}{2m}, \frac{1}{2l}$ for $y_i = 0,1$ respectively, where $m$ and $l$ are the number of negatives and positives respectively.
  - For $t = 1, \ldots, T$ :
    1. Normalize the weights, $w_{t,i} \leftarrow \frac{w_{t,i}}{\sum_{j-1}^{n} w_{t,i}}$ so that $w_t$ is a probability distribution.
    2. Send the weights to all slaves.
    3. On each slave:
        a. For each feature, $j$, train a classifier $h_j$ which is restricted to using a single feature. The error is evaluated with respect to $w_t$, $\epsilon_j = \sum_i w_i |h_j(x_i) - y_i|$.
        b. Send the classifier, $h_t$, with the lowest error $\epsilon_t$, to master workstation.
    4. On master workstation:
        a. Amongst the received classifiers from each slave, choose the classifier, $h_t$, with the lowest error $\epsilon_t$.
        b. Update the weights: $w_{t+1,i} = w_{t,i} \beta_t^{1-e_i}$ where $e_i = 0$ if example $x_i$ is classified correctly, $e_i = 1$ otherwise, and $\beta_t = \frac{\epsilon_t}{1-\epsilon_t}$.
- The final strong classifier is:

$$h(x) = \begin{cases} 1, & \sum_{t=1}^{T} \alpha_t h_t(x) \geq \frac{1}{2}\sum_{t=1}^{T} \alpha_t \\ 0, & otherwise \end{cases}$$

where $\alpha_t = \log\frac{1}{\beta_t}$

### 3.3.3. Web Services and Parallel Execution on two Hierarchal Levels

The previous technique divided the work based on feature type. Now we further distribute the calculations in a feature type to another set of machines in the next hierarchical level as shown in figure 5.

**Pseudocode of Two level Master, five sub-master, and N slaves**

- Given example images $(x_1, y_1), \ldots, (x_n, y_n)$ where $y_i = 0,1$ for negative and positive examples respectively.
- Prepare one master workstation, five sub-masters, and twenty five slaves.
- Each sub-master and its slaves are assigned to one particular feature type.
  - (sub-master 1 and 5 slaves, Three rectangles Horizontal)
  - (sub-master 2 and 5 slaves, Three rectangles Vertical)
  - (sub-master 3 and 5 slaves, Two rectangles Horizontal)
  - (sub-master 4 and 5 slaves, Two rectangles Vertical)
  - (sub-master 5 and 5 slaves, Four rectangles)
- On all slaves' workstations: Initialize all images on each slave.
- On master workstation:
  - Initialize weights $w_{1,i} = \frac{1}{2m}, \frac{1}{2l}$ for $y_i = 0,1$ respectively, where m and l are the number of negatives and positives respectively.
  - For $t = 1, \ldots, T$ :
    1. Normalize the weights, $w_{t,i} \leftarrow \frac{w_{t,i}}{\sum_{j-1}^{n} w_{t,i}}$ so that $w_t$ is a probability distribution.
    2. Send the weights to all sub-masters, and then sub-master sends them to its slaves.
    3. Sub-masters divide the features between their slaves, where each slave responsible for some parts of the features.
    4. On each slave:
       a. For each feature,j, train a classifier $h_j$ which is restricted to using a single feature. The error is evaluated with respect to $w_t$, $\epsilon_j = \sum_i w_i |h_j(x_i) - y_i|$.
       b. Send the classifier,$h_t$, with the lowest error $\epsilon_t$, to the assigned sub-master workstation.
       c. Each sub-master choose the classifier,$h_t$, with the lowest error $\epsilon_t$ amongst their slaves and send it to the master.
    5. On master workstation:
       a. Amongst the received classifiers from each sub-master, choose the classifier,$h_t$, with the lowest error $\epsilon_t$.
       b. Update the weights: $w_{t+1,i} = w_{t,i}\beta_t^{1-e_i}$ where $e_i = 0$ if example $x_i$ is classified correctly, $e_i = 1$ otherwise, and $\beta_t = \frac{\epsilon_t}{1-\epsilon_t}$.
- The final strong classifier is:

$$h(x) = \begin{cases} 1, & \sum_{t=1}^{T} \alpha_t h_t(x) \geq \frac{1}{2} \sum_{t=1}^{T} \alpha_t \\ 0, & \text{otherwise} \end{cases}$$

where $\alpha_t = \log \frac{1}{\beta_t}$

## 3. EXPERIMENTAL RESULTS

We developed four variations of the AdaBoost algorithm, as follows:

- Sequential algorithm.
- Parallel on one machine that only uses TPL.
- Web Services and Parallel execution on one hierarchal level.
- Web Services and Parallel execution on two hierarchal levels.

Table 3 shows a comparison of the different approaches we implemented. These results show a significant improvement in speedup as compared to the previous work reported in [1]. We obtain a speedup of 95.1 as compared to a speedup of 2.6 reported in [1]. Figure 6 shows the parallel execution time of our implementation as the number of slaves is increased. With 31 machines, an execution time per feature of 4.8 second is achieved.

Table 3. Comparison of all used approaches (Times are in seconds).

| | Uploading time to memory – done one time only (seconds) | Average execution time for each round (seconds) | Speed-up (with respect to sequential execution) |
|---|---|---|---|
| **Sequential alg. On one PC** | 1780.6 | 456.5 | ---- |
| **Parallel alg. on one PC** | 330.7 | 116.1 | 3.9 |
| **Parallel and distributed one-level architecture on 6 PC's** | 92.7 | 24.6 | 18.6 |
| **Parallel and distributed two-level architecture on 21 PC's** | 30.3 | 6.4 | 71.3 |
| **Parallel and distributed two-level architecture on 26 PC's** | 35.4 | 5.2 | 87.8 |
| **Parallel and distributed two-level architecture on 31 PC's** | 31.7 | 4.8 | 95.1 |

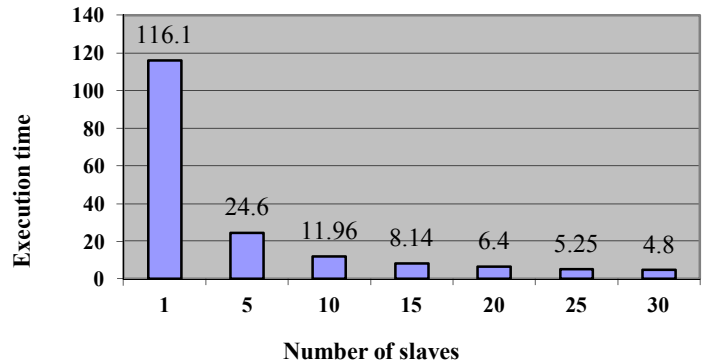

Figure 6. Parallel execution time based on total number of slaves workstations

To be able to predict the speedup for any number of machines available, we develop the following predictive equation for calculating parallel execution time based on the number of nodes in the last level attached to one sub-master node in the middle level (see figure 6 and 7).

$$Parallel\ execution = (0.2 * n) + \left(\frac{0.5}{1000}\right) * \left(\frac{m}{n}\right)$$

Where $n$ is the number of nodes attached to a one sub-master node, and $m$ is the maximum number of features allocated to one sub-master node.

It is noticed that increasing number of nodes in the last level beyond 7 per feature type is not going to further help in speeding up execution. Since communication overhead in the network is going to dominate. Table 4, 5 and 6, and figure 6 explain that.

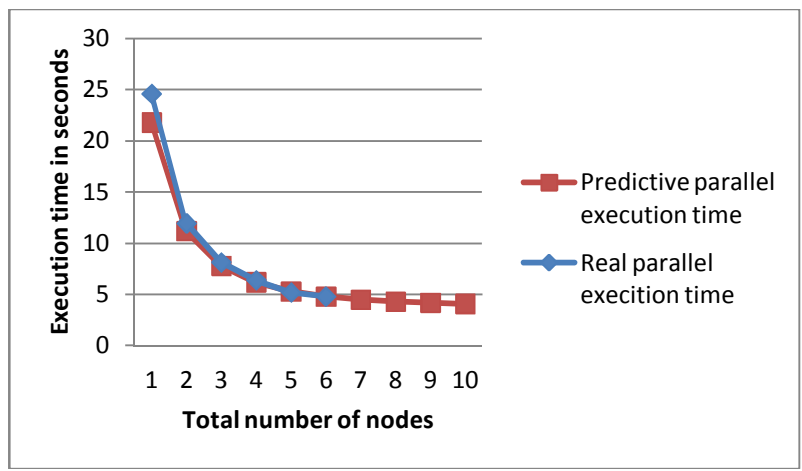

Figure 7. Real and predictive parallel execution time based on total number of slaves workstations in the last level

Table 4. The result of the predictive equation based on number of nodes attached to one sub-master node.

| Number of nodes | Execution time per round (seconds) |
|---|---|
| 1 | 21.8 |
| 2 | 11.2 |
| 3 | 7.8 |
| 4 | 6.2 |
| 5 | 5.3 |
| 6 | 4.8 |
| 7 | 4.5 |
| 8 | 4.3 |
| 9 | 4.2 |
| 10 | 4.1 |

Table 5. Overhead on the one level network using master and 5 slaves only.

| | Average overhead on network per round (mille seconds) |
|---|---|
| 4 rectangle node | 251.04 |
| 3 rectangle vertical node | 257.8 |
| 3 rectangle horizontal node | 384.8 |
| 2 rectangle vertical node | 253.3 |
| 2 rectangle horizontal node | 356.61 |

Table 6. Overhead on the two levels network using master, 5 sub-master, and 25 slaves.

| | Average overhead on network per round (mille seconds) |
|---|---|
| 4 rectangle node | 280.2 |
| 3 rectangle vertical node | 283.43 |
| 3 rectangle horizontal node | 334.82 |
| 2 rectangle vertical node | 294.86 |
| 2 rectangle horizontal node | 410.3 |

## 5. CONCLUSIONS

We have developed a hybrid parallel and distributed implementation of AdaBoost algorithm that exploits the multiple cores in a CPU via light weight threads, and also uses multiple machines via web service software architecture to achieve high scalability. We also develop a novel hierarchical web services based distributed architecture for maximal exploitation of concurrency in the AdaBoost algorithm. We demonstrate nearly linear speedup upto the number of processors available to us, and can accomplish the learning of a feature in the AdaBoost algorithm within a few seconds. This may be particularly useful in applications where the classifier needs to be dynamically adapted to changing training set data e.g., in car model detection. In comparison with the previously published work, which used a single level master slave parallel and distributed implementation [8], and only achieved a speedup of 2.66 using four nodes, we achieve a speedup of 95.1 on 31 workstations each having a quadcore processor, resulting in a learning time of only 4.8 seconds per feature. Our future work involves developing a learning framework that can adapt to different object detection needs in a near real time manner.